\documentclass[12pt]{article}
\usepackage{graphicx}
\def \app{D_{\pi \pi}}
\def \b{{\cal B}}

\def \bea{\begin{eqnarray}}
\def \beq{\begin{equation}}
\def \bl{\bar\lambda}
\def \bo{B^0}

\def \cn{Collaboration}
\def \cpp{C_{\pi \pi}}
\def \eea{\end{eqnarray}}
\def \eeq{\end{equation}}
\def \ite{{\it et al.}}

\def \lpp{\lambda_{\pi \pi}}
\def \ob{\overline{B}^0}
\def \ok{\overline{K}^0}
\def \rpp{R_{\pi \pi}}

\def \spp{S_{\pi \pi}}
\begin{document}
\begin{flushright}
TECHNION-PH-2004-21\\
CLNS-04-1879 \\
hep-ph/0405173 \\
May 2004 \\
\end{flushright}
\renewcommand{\thesection}{\Roman{section}}
\renewcommand{\thetable}{\Roman{table}}
\centerline{\bf IMPLICATIONS OF CP ASYMMETRIES IN $B \to \pi^+ \pi^-$
\footnote{To be submitted to Physics Letters B.}}
\medskip
\centerline{Michael Gronau}
\centerline{\it Physics Department, Technion -- Israel Institute of Technology}
\centerline{\it 32000 Haifa, Israel}
\medskip
\centerline{Jonathan L. Rosner\footnote{On leave from
Enrico Fermi Institute and Department of Physics, University of Chicago,
Chicago, Illinois 60637.}}
\centerline{\it Laboratory for Elementary-Particle Physics, Cornell University}
\centerline{\it Ithaca, New York 14850}
\bigskip
\begin{quote}

CP asymmetries in $B^0(t)\to \pi^+ \pi^-$ are studied by relating this process
in broken flavor SU(3) with  $B^+\to K^0\pi^+$ and $B^0\to K^+\pi^-$.  Using
two different scenarios for SU(3) breaking, we show that the range of values of
the weak phase $\alpha$ permitted by the measured asymmetries overlaps with
that obtained from other CKM constraints, supporting the KM origin of the
asymmetries.  We evaluate the potential precision of this method to improve the
determination of $\alpha$.
\end{quote}

\leftline{\qquad PACS codes:  12.15.Hh, 12.15.Ji, 13.25.Hw, 14.40.Nd}

\bigskip
Measurements of CP-violating asymmetries in the decays $B^0(t) \to \pi^+ \pi^-$
and their charge conjugates have reached a very interesting stage.  The 
BaBar~\cite{BaBar} and Belle~\cite{Belle} collaborations measured two 
asymmetries, $\cpp\equiv -A_{\pi\pi}$ and $\spp$,  defined by~\cite{MG}
\beq
\frac {\Gamma(\ob(t) \to \pi^+\pi^-) - \Gamma(\bo(t)\to\pi^+\pi^-)}
 {\Gamma(\ob(t) \to \pi^+\pi^-) - \Gamma(\bo(t)\to\pi^+\pi^-)} =
 -\cpp\cos(\Delta mt) + \spp\sin(\Delta mt)~,
 \eeq
 obtaining values
 \beq\label{BaBe}
\cpp = \left\{ \begin{array}{c} -0.19 \pm 0.19 \pm 0.05~, \cr
-0.58 \pm 0.15 \pm 0.07~,\end{array} \right.
\spp = \left\{ \begin{array}{c} -0.40 \pm 0.22 \pm 0.03~,~~~~
{\rm BaBar}~, \cr
-1.00 \pm 0.21 \pm 0.07~,~~~~
{\rm Belle}~.\end{array} \right.
\eeq
The Belle measurement rules out the case of CP-conservation, $\cpp=\spp=0$,
at a level of 5.2 standard deviations.  The current average values of the two
asymmetries are~\cite{HFAG}
\beq\label{CSave}
\cpp = -0.46 \pm 0.13~,~~~\spp = -0.74 \pm 0.16~.
\eeq

An immediate question, motivated by a search for physics beyond the 
Standard Model,  is whether these values are consistent with other constraints 
on Cabibbo-Kobayashi-Maskawa (CKM) parameters.
If confirmed, the next  question is whether reducing experimental 
errors in the asymmetries may improve these constraints, thereby 
tightening the current range of the weak phase 
$\alpha \equiv \phi_2$~\cite{CKMfitter}, $75^\circ < \alpha < 120^\circ$.

An extraction of $\alpha$ from the CP asymmetry in $B^0\to\pi^+\pi^-$ 
is obstructed by the effect of a penguin amplitude~\cite{MG,LP}.
The theoretically cleanest way of obtaining  
$\alpha$ from these measurements is based on isospin 
symmetry~\cite{iso}. It includes electroweak penguin effects~\cite{EWP}, 
and requires in addition to the measured rate of $B^+\to \pi^+\pi^0$ separate 
decay rate measurements of $\bo$ and $\ob$ to $\pi^0\pi^0$. Isospin breaking 
effects are expected to introduce an uncertainty of only a few degrees in the 
determination of the weak phase. Prior to a $\bo/\ob$ separation, the measured
combined decay rate of $\bo$ and $\ob$ into $\pi^0\pi^0$ provides a measure
for the uncertainty in $\alpha$~\cite{GQ}.  Current branching ratios imply an 
uncertainty of about $50^\circ$ for arbitrary asymmetry 
measurements~\cite{BaBar,FPCP03}. A higher precision may be achieved 
for special values of  the asymmetries~\cite{BSBS,Ali}.
In order to obtain more precise knowledge of $\alpha$
before $\bo\to\pi^0\pi^0$ and $\ob\to\pi^0\pi^0$ are separately measured, 
further assumptions beyond isospin symmetry are required.

A powerful approach to $B$ decays into a pair of charmless pseudoscalar 
mesons is based on the broader but the less precise flavor SU(3) 
symmetry~\cite{SU3,GHLR}. Introducing SU(3) breaking effects in a 
controllable and testable manner~\cite{SU3br} improves the precision of 
this approach. A variety of studies along this line, focusing 
on $B\to \pi\pi$ and $B\to K\pi$ decays, were performed in the past ten 
years~\cite{refs}. A crucial factor in determining $\alpha$ in $B^0\to \pi^+\pi^-$
is a knowledge of the ratio of penguin and tree amplitudes contributing to 
this process~\cite{GR02}.

In the present Letter we update and modify an analysis~\cite{Pcon} of
$B^0(t)\to\pi^+\pi^-$, which combines this process with $B^+\to K^0\pi^+$. 
In~\cite{Pcon} we assumed that both tree and penguin amplitudes factorize. 
Instead, we will now leave open the question of factorization of penguin 
amplitudes, comparing results obtained under different assumptions about SU(3)
breaking in these amplitudes. Using current data unavailable at the time of the
analysis in \cite{Pcon}, we will argue for a  ratio of penguin-to-tree 
amplitudes $P/T$ larger than commonly accepted. This study will also be
combined with a complementary analysis relating $B^0\to \pi^+\pi^-$ and $B^0
\to K^+\pi^-$, where similar bounds on $P/T$ are obtained. Earlier but somewhat
different studies relating these two last processes were performed in
\cite{SW,DHGR,ChFl}.  Finally, we use information on $P/T$ from $K\pi$ and
$\pi\pi$ rates to study the CP asymmetries in $B\to \pi^+\pi^-$ as functions of
$\alpha$.  We will show that the current asymmetries are consistent with the
allowed range of $\alpha$, and will discuss the possibility of tightening this
range.

We use the ``c-convention"~\cite{Pcon}, in which the top-quark has been
integrated out in the $b\to d$ penguin transition and unitarity of the CKM
matrix has  been used. Absorbing a $P_{tu}$ term in $T$, one writes 
\beq\label{Apipi}
A(B^0\to\pi^+\pi^-) = Te^{i\gamma} + P e^{i\delta}~.
\eeq
By convention $T$ and $P$, which involve magnitudes of CKM factors, 
$|V^*_{ub}V_{ud}|$ and $|V^*_{cb}V_{cd}|$, are positive and the strong 
phase $\delta$ lies in the range $-\pi \le \delta \le \pi$. The amplitude for 
$\ob \to\pi^+\pi^-$ is obtained by changing the sign of $\gamma$.
The asymmetries $\cpp$ and $\spp$ are given by~\cite{MG}
\beq \label{eqn:CSpipi}
\cpp \equiv \frac{1 - |\lpp|^2}{1 + |\lpp|^2}~~,
~~~~\spp \equiv \frac{2 {\rm Im}(\lpp)}{1 + |\lpp| ^2}~~,
\eeq
where
\beq
\lpp \equiv e^{-2i \beta} \frac{A(\ob \to \pi^+ \pi^-)}
{A(B^0 \to \pi^+ \pi^-)}~~~.
\eeq
Substituting (\ref{Apipi}) into these definitions, one obtains~\cite{GR02},
\bea\label{C}
\cpp & = & \frac{2r\sin\delta\sin(\beta +\alpha)}{\rpp}~,
\\
\label{S}
\spp & = & \frac{\sin 2\alpha + 2r\cos\delta\sin(\beta-\alpha) - 
r^2\sin 2\beta}{\rpp}~,
\\
\label{Rpp}
\rpp & = & 1 - 2r\cos\delta\cos(\beta + \alpha) + r^2~,
\eea
where
\beq
r \equiv \frac{P}{T}~,
\eeq
is a ratio of penguin to tree amplitudes.

In the absence of a penguin amplitude ($r=0$) one has 
$\cpp=0,~\spp=\sin 2\alpha$. For small values of $r$, keeping only linear terms 
in this ratio, one finds
\bea
\cpp & = & 2r\sin\delta\sin(\beta + \alpha) + {\cal O}(r^2)~,
\\
\spp & = & \sin 2\alpha + 2r\cos\delta\sin(\beta + \alpha)\cos 2\alpha
+ {\cal O}(r^2)~.
\eea
That is, in the linear approximation the allowed region in the $(\spp,\cpp)$
plane is confined to an ellipse centered at $(\sin 2 \alpha, 0)$,
with semi-principal axes $2[r\sin(\beta + \alpha)\cos 2\alpha]_{\rm max}$
and $2[r\sin(\beta + \alpha)]_{\rm max}$.  In our study 
below we will use the exact expressions (\ref{C})--(\ref{Rpp}).

Given a value of $\beta$, as already measured in $B^0(t) \to J/\psi K_S$
\cite{psiKs}, the two measurables $\cpp$ and $\spp$ 
provide two equations for
the weak phase $\alpha$ and for the two hadronic parameters $r$ and $\delta$.
At least one additional constraint on $r$ and $\delta$ is needed in order to
determine $\alpha$.  Such constraints are given (by isospin and) by flavor
SU(3) symmetry considerations as described below.

Turning now to $B\to K\pi$ decays, one describes corresponding decay amplitudes
in terms of primed quantities, $T'$ and $P'$~\cite{GHLR}.  We introduce an 
SU(3) breaking factor $f_K/f_\pi$ in tree amplitudes which are expected to 
factorize~\cite{BBNS,SCET}, but assume in the first place exact SU(3) for
penguin amplitudes for which factorization is not 
expected to hold~\cite{charmingP},
\beq\label{T'P'}
T' = \frac{f_K}{f_\pi}\frac{V^*_{ub}V_{us}}{V^*_{ub}V_{ud}}\,T = 
\frac{f_K}{f_\pi}\bl\,T~,~~~~P' = \frac{V^*_{cb}V_{cs}}{V^*_{cb}V_{cd}}\,P
= -\bl^{-1}P~.
\eeq
Here 
\beq
\bl \equiv \frac{\lambda}{1 - \lambda^2/2} = 0.230~.
\eeq
The assumption of SU(3) symmetry in penguin amplitudes can be
tested~\cite{Uspin} by comparing the measured rate of $B^+\to K^0\pi^+$
with future measurements of $B^+\to K^+\ok$. Another test of this assumption
and the effect of possible SU(3) breaking in $P$ will be discussed below. SU(3)
amplitudes represented by exchange and annihilation contributions
occur in $B^0\to\pi^+\pi^-$ and $B^+ \to K^0\pi^+$ respectively~\cite{GHLR}.
They  are $1/m_b$ suppressed relative to tree  and penguin
amplitudes~\cite{SCET} and will be neglected. These approximations and the
neglect of very small color-suppressed electroweak penguin contributions are
testable in $B^0 \to K^+K^-$ and in other processes~\cite{rescat}.

Under these assumptions one may write expressions for $B\to K\pi$
amplitudes in terms of amplitudes contributing to $B^0 \to \pi^+\pi^-$:
\bea\label{AKpi+}
A(B^+ \to K^0\pi^+) & = & -\bl^{-1}Pe^{i\delta}~,
\\
\label{AKpi-}
A(B^0 \to K^+\pi^-) & = & - \frac{f_K}{f_\pi}\bl\,Te^{i\gamma} + 
\bl^{-1}Pe^{i\delta}~.
\eea
The CP asymmetry in the first process vanishes, while that of $B^0\to K^+\pi^-$
is related to the asymmetry in $B^0\to \pi^+\pi^-$~\cite{DHGR},
\beq\label{DeltaGamma}
\Gamma(\ob \to K^-\pi^+) - \Gamma(\bo\to K^+\pi^-) = - \frac{f_K}{f_\pi}
[\Gamma(\ob \to \pi^+\pi^-) - \Gamma(\bo\to \pi^+\pi^-)]~.
\eeq
Here and below we neglect phase space factors introducing calculable 
corrections at a percent level. Eq.~(\ref{DeltaGamma})  may be used to test  
SU(3) symmetry including the SU(3) breaking factor $f_K/f_\pi$.
This equality reads in units of $10^6$ times branching ratios
\beq\label{Delta}
-3.5 \pm 1.0 = - 5.2 \pm 1.5~~~,
\eeq
where we use the current charge-averaged branching ratios, in units of
$10^{-6}$ \cite{HFAG}:
\beq\label{BR}
\bar\b(B\to \pi^+\pi^-) = 4.6 \pm 0.4,~
\bar\b(B\to K^0\pi^+) = 21.8 \pm 1.4,~
\bar\b(B\to K^+\pi^-) = 18.2 \pm 0.8~,
\eeq
and the CP asymmetry \cite{HFAG} $A(K^+ \pi^-) = -0.095 \pm 0.028$ and the
average (3) for $\cpp$.  Although current errors are too large to provide a
quantitative test of flavor SU(3),
the consistency of the signs of the two asymmetries provides one test of SU(3).
With future increased statistics, Eq.~(\ref{DeltaGamma}) may be used as an
additional input in a determination of $\alpha$ as described below. 

Each of the two charge averaged rates $\bar\Gamma(B^+\to K^0\pi^+)\equiv
[\Gamma(B^+\to K^0\pi^+)+\Gamma(B^-\to \ok \pi^-)]/2$ and $\bar\Gamma(B^0 \to
K^+\pi^-)\equiv [\Gamma(\bo\to K^+\pi^-) +\Gamma(\ob\to K^-\pi^+)]/2$ provides
an additional constraint on the three parameters $r,~\delta$ and $\alpha$.  We
normalize these rates by the charge averaged rate of decays to $\pi^+\pi^-$,
$\bar\Gamma(B^0\to\pi^+\pi^-) \equiv [\Gamma(\bo\to \pi^+\pi^-) + \Gamma(\ob\to
\pi^+\pi^-)]/2$, defining the following two ratios (${\cal R}_+$ corresponds to
$1/b_c$ defined in~\cite{Pcon}):
\bea
{\cal R}_+ & \equiv & \frac{\bl^2\,\bar\Gamma(B^+\to K^0\pi^+)}
{\bar\Gamma(B^0\to \pi^+\pi^-)}~,
\\
{\cal R}_0 & \equiv & \frac{\bl^2\,\bar\Gamma(B^0 \to K^+\pi^-)}
{\bar\Gamma(B^0\to \pi^+\pi^-)}~.
\eea
The values (\ref{BR}) and the lifetime ratio \cite{tau} $\tau(B^+)/\tau(B^0) =
1.077 \pm 0.013$ imply
\beq\label{R+,0}
{\cal R}_+ = 0.235 \pm 0.026~,~~~~{\cal R}_0 = 0.209 \pm 0.020~,
\eeq
in which the errors are already at a level of only 10$\%$.

Substituting Eqs.~(\ref{Apipi}), (\ref{AKpi+}) and (\ref{AKpi-}), we obtain
\bea\label{R+r}
{\cal R}_+ & = & \frac{r^2}{\rpp}~,
\\
\label{R0r}
{\cal R}_0 & = & \frac{r^2 + 2r\bl'^2\cos\delta\cos(\beta +\alpha) 
+ \bl'^4}{\rpp}~,~~~~\bl' \equiv \sqrt{\frac{f_K}{f_\pi}}\,\bl~.
\eea
These expressions and Eq.\ (\ref{Rpp}) can be inverted to write $r$ in terms of 
$z \equiv \cos\delta\cos(\beta + \alpha)=-\cos\delta\cos\gamma$ and 
one of the two measured quantities ${\cal R}_+$ or ${\cal R}_0$:
\bea\label{rR+}
{\cal R}_+:~~~~~r & = & \frac{\sqrt{{\cal R}_+^2\,z^2 + (1 - {\cal R}_+)
{\cal R}_+} - {\cal R}_+\,z}{1 - {\cal R}_+}~,
\\
\label{rR0}
{\cal R}_0:~~~~~r & = & \frac{\sqrt{({\cal R}_0 +\bl'^2)^2\,z^2 +
(1 - {\cal R}_0)({\cal R}_0 - \bl'^4)} - ({\cal R}_0 + \bl'^2)\,z}
{1 - {\cal R}_0}~.
\eea
In principle, Eqs.\ (\ref{Rpp}), (\ref{R+r}), and (\ref{R0r}) may be solved for
$r$ (and $z$) in terms of ${\cal R}_+$ and ${\cal R}_0$,
\beq
r = \sqrt{\frac{{\cal R}_+(1 + \bl'^2)}{1- {\cal R}_+ 
+\bl'^{-2}({\cal R}_0 - {\cal R}_+)}}~.
\eeq
However, in practice this provides no useful information about $r$ because 
small errors in ${\cal R}_+$ and ${\cal R}_0$ are enhanced by the factor 
$\bl'^{-2}$ multiplying ${\cal R}_0 - {\cal R}_+$ in the denominator, thereby 
permitting very large values of $r$.

At this point, let us consider lower and upper bounds on $r$
following separately from Eqs.~(\ref{rR+}) and (\ref{rR0}),
depending on branching ratio measurements of $B^+\to K^0\pi^+$
and $B^0\to K^+\pi^-$, respectively.
For the values of ${\cal R}_+$ and ${\cal R}_0$ in (\ref{R+,0}),
both expressions for $r$ are monotonically decreasing functions of $z$. 
Using current constraints on CKM parameters~\cite{CKMfitter} implying
$38^\circ \le \gamma \le 80^\circ$ at $95\%$ confidence level, the lowest
and highest allowed value of $z$ are --0.79 and 0.79, respectively. 
Inserting these values in (\ref{rR+}) and (\ref{rR0}), and using central values 
in (\ref{R+,0}), we find the following bounds:
\bea\label{R+}
{\cal R}_+:~~~~~0.36 \le &  r & \le 0.85~,
\\
\label{R0}
{\cal R}_0:~~~~~0.30 \le & r & \le 0.85~.
\eea
Slightly wider ranges are allowed when including errors in 
${\cal R}_+$ and ${\cal R}_0$.

Values of $r$ in the lower parts of these ranges 
correspond to $z > 0$ or $\pi/2 < |\delta| < \pi$,  while the upper parts  
correspond to $z < 0$ or $0 < |\delta| < \pi/2$. Assuming that $\delta$ 
lies in the first (positive or negative) quadrant, $|\delta| < \pi/2$, one has
\bea \label{R+del}
{\cal R}_+:~~~~~0.55 \le &  r & \le 0.85~~~~({\rm assuming}~|\delta| < \pi/2)~,
\\
\label{R0del}
{\cal R}_0:~~~~~0.51 \le & r & \le 0.85~~~~({\rm assuming}~|\delta| < \pi/2)~.
\eea
The two lower bounds, which become slightly smaller (0.51 and 0.48 
respectively) when errors are included, stand in contrast to most
calculations based on QCD factorization ($r=0.285 \pm 0.076$~\cite{BBNS} 
or $r=0.32^{+0.16}_{-0.09}$~\cite{BN}) and perturbative QCD 
($r=0.23^{+0.07}_{-0.05}$~\cite{KS}), where values of $\delta$ were obtained
in the range $|\delta| < \pi/2$. 
A value $r=0.26\pm 0.08$ was estimated~\cite{LR} by applying 
factorization to $B\to \pi\ell\nu$, but disregarding the $P_{tu}$ term in 
$T$. On the other hand, a recent a global SU(3) fit to all $B\to\pi\pi$ 
and $B\to K\pi$ decays ~\cite{BPP}, including a sizable $P_{tu}$ 
contribution, obtained values $r = 0.69 \pm 0.09$ and
$\delta = (-34^{+11}_{-25})^\circ$, in obvious agreement with the bounds 
(\ref{R+del}) and ({\ref{R0del}). Note that these bounds do not rely on 
the asymmetry measurements in $B^0\to \pi^+\pi^-$. As we will see below, 
where we make no assumption about $\delta$, the measured asymmetries also seem
to favor negative values of $\delta$ in the first quadrant.

Each of the two relations (\ref{rR+}) and (\ref{rR0}) may be used separately
together with (\ref{C})--(\ref{Rpp}) to express $\cpp$ and $\spp$ in terms of 
$\delta,~\alpha$ and the measured values of $\beta, {\cal R}_+$ or
${\cal R}_0$.
\begin{figure}
\includegraphics[height=4.9in]{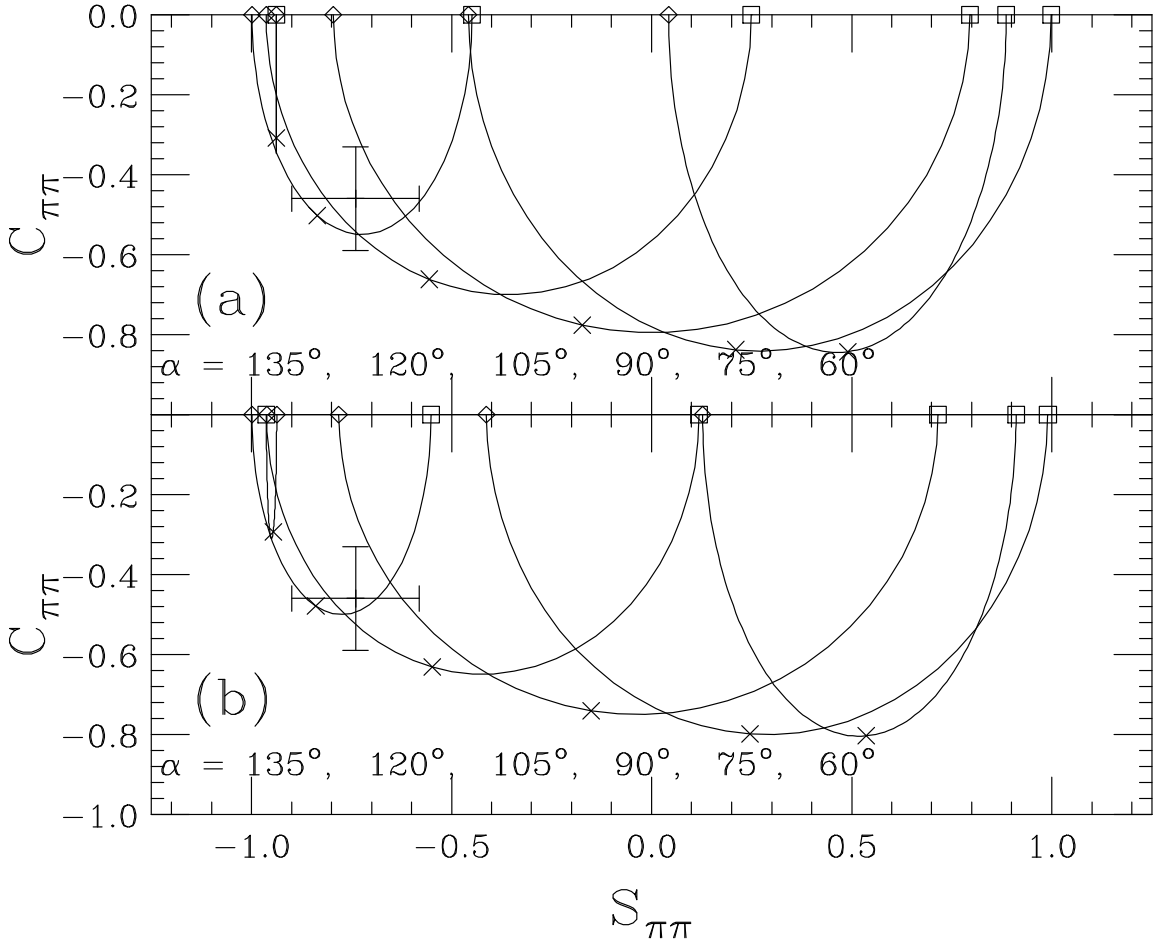}
\caption{Parametric curves of $C_{\pi \pi}$ vs.\ $S_{\pi \pi}$ for the
range $-\pi \le \delta \le 0$, based on measurements of (a) ${\cal R}_+$ and
(b) ${\cal R}_0$.  Points marked with diamonds, crosses and squares denote
$\delta = 0$, $-\pi/2$ and $-\pi$, respectively.  Plotted point denotes the
experimental average (3).}
\end{figure}
We draw two separate plots, using in one case the measurement  
of  ${\cal R}_+$ and in the other case that of ${\cal R}_0$. Values of $\spp$
and $\cpp$, for $\beta = 23.7^\circ$~\cite{CKMfitter}, the central value of
${\cal R}_+$ in (\ref{R+,0}), and for a set of four values of $\alpha$ in the
currently allowed range~\cite{CKMfitter}~$75^\circ \le \alpha \le 120^\circ$
and two values outside this range,
are plotted in Fig.~1(a).  We plot only the case $\delta \le 0$ since the
experimental average of BaBar and Belle values corresponds to $C_{\pi \pi}
\le 0$, and the signs of $\sin \delta$ and $C_{\pi \pi}$ are correlated
by Eq.\ (11).

As anticipated, curves of fixed $\alpha$ and varying $\delta$ are approximate
ellipses. Points marked with diamonds, crosses and squares denote $\delta = 0$,
$-\pi/2$ and $-\pi$, respectively.  A consequence of the second term in the
numerator of (\ref{S}) is that a point $\delta = -\pi$ on each ellipse is
located to the right of a point $\delta = 0$ on the same ellipse.  The plotted
point including errors corresponds to the present averaged asymmetries
(\ref{CSave}). Fig.~1(b) is plotted in an analogous manner, using the central
value of ${\cal R}_0$ in (\ref{R+,0}).

We note the similar dependence in Figs.~1(a) and 1(b) of $\spp$ and $\cpp$ as
functions of $\delta$ and $\alpha$.  This common behavior supports our
assumption of flavor SU(3), also adding to the statistical significance of the
plots, which are based on central values of measurements of $B^+\to K^0\pi^+$
and $B^0 \to K^+\pi^-$.  The approximate ellipses in Fig.~1(a), for fixed
values of $\alpha$ and varying $\delta$, are only slightly larger than those in
Fig.~1(b). This follows from the somewhat larger values of $r$ permitted by
${\cal R}_+$ than those allowed by ${\cal R}_0$, as given in the bounds
(\ref{R+}) and (\ref{R0}). 

An important question is: what can be learned about $\delta$ and $\alpha$ from
the present average asymmetries (\ref{CSave})?  A negative value of $\cpp$,
favored by the data, implies $-\pi < \delta < 0$.  A negative $\spp$, supported
by both the BaBar and the Belle results (\ref{BaBe}), favors $-\pi/2 < \delta
< 0$ for all plotted values of $\alpha$ except $\alpha = 120^\circ$ and
$135^\circ$.  

As for $\alpha$, it is already remarkable that the two measured asymmetries 
lie in an area in the $(\spp,\cpp)$ plane overlapping with that corresponding 
to the range $75^\circ < \alpha < 120^\circ$ obtained from other constraints
\cite{CKMfitter}.  Larger values in this range are favored.  The measured
asymmetries exclude smaller values of $\alpha$ than in this range (e.g., 
$\alpha = 60^\circ$), for which corresponding ellipses lie too much to the
right (e.g., corresponding to $\spp \ge 0$).
Larger values of $\alpha$ than in this range (e.g., $\alpha = 135^\circ$) are
described by shorter and narrower ellipses, implying values of $|\cpp|$ smaller than the
measured central value.  The consistency between the range of $\alpha$ allowed
by the asymmetries and by all other constraints  is certainly nontrivial,
indicating that the origin of the asymmetry is largely the KM phase.

Using Fig.~1(b) [slightly more restrictive than Fig.~1(a)], an error ellipse
with center and principal axes specified as in Eq.\ (3) just touches the curve
for $\alpha = 86^\circ$ at a single point, excluding lower values and implying
$\alpha = (103 \pm 17)^\circ$.  An actual determination of $\alpha$ at this
precision (rather than the use of the upper bound of $120^\circ$ as we have
done) requires reducing the experimental errors in the two asymmetries, since
as $\alpha$ grows one must contend with discrete ambiguities in which curves
for different $\alpha$ intersect at a single point.  Neglecting for the moment
this feature, the horizontal distance between the two ellipses drawn for
$\alpha=90^\circ$ and $105^\circ$ corresponds to a change in $\spp$ of
magnitude $\Delta\spp= 0.18$.  A reduction of the current experimental error,
$\Delta\spp = 0.16$, by a factor two will result in a comparable reduction in
the error of $\alpha$ to $\Delta\alpha = 9^\circ$.  This seems like an ultimate
precision considering the approximations made in this analysis.  Both Figs.\
1(a) and 1(b) show that as $\alpha$ approaches its current upper limit of
$120^\circ$ a
higher precision in $\spp$ is required in order to achieve this precision in
$\alpha$, both because of the discrete ambiguity just mentioned and because
the curves for a given change in $\alpha$ lie closer to to one another.

Before concluding, we wish to comment on the effect of SU(3) breaking in $P$. 
For illustration, let us assume $P' = -(f_K/f_\pi)\bl^{-1}P$ instead of
(\ref{T'P'}), as would be the case if penguin amplitudes were to factorize.
In this case the right-hand-side of (\ref{DeltaGamma}) includes
a factor $(f_K/f_\pi)^2$, implying a central value on the right-hand-side of
(\ref{Delta}),
$-6.3 \pm 1.9$,
 almost twice as large as the central value on
the left-hand-side.  As a result, one must replace ${\cal R}_+ \to {\cal R}_+/
(f_K/f_\pi)^2$ in (\ref{rR+}), and ${\cal R}_0 \to {\cal R}_0/(f_K/f_\pi)^2,
~\bl' \to \bl$ in (\ref{rR0}).  The bounds (\ref{R+del}) and (\ref{R0del}) are
replaced by tighter ones,
$0.43 \le r \le 0.61$
using ${\cal R}_+$, and
$0.40 \le r \le 0.61$
using ${\cal R}_0$.
The two lower bounds are still somewhat higher than most QCD calculations. 

\begin{figure}
\includegraphics[height=4.9in]{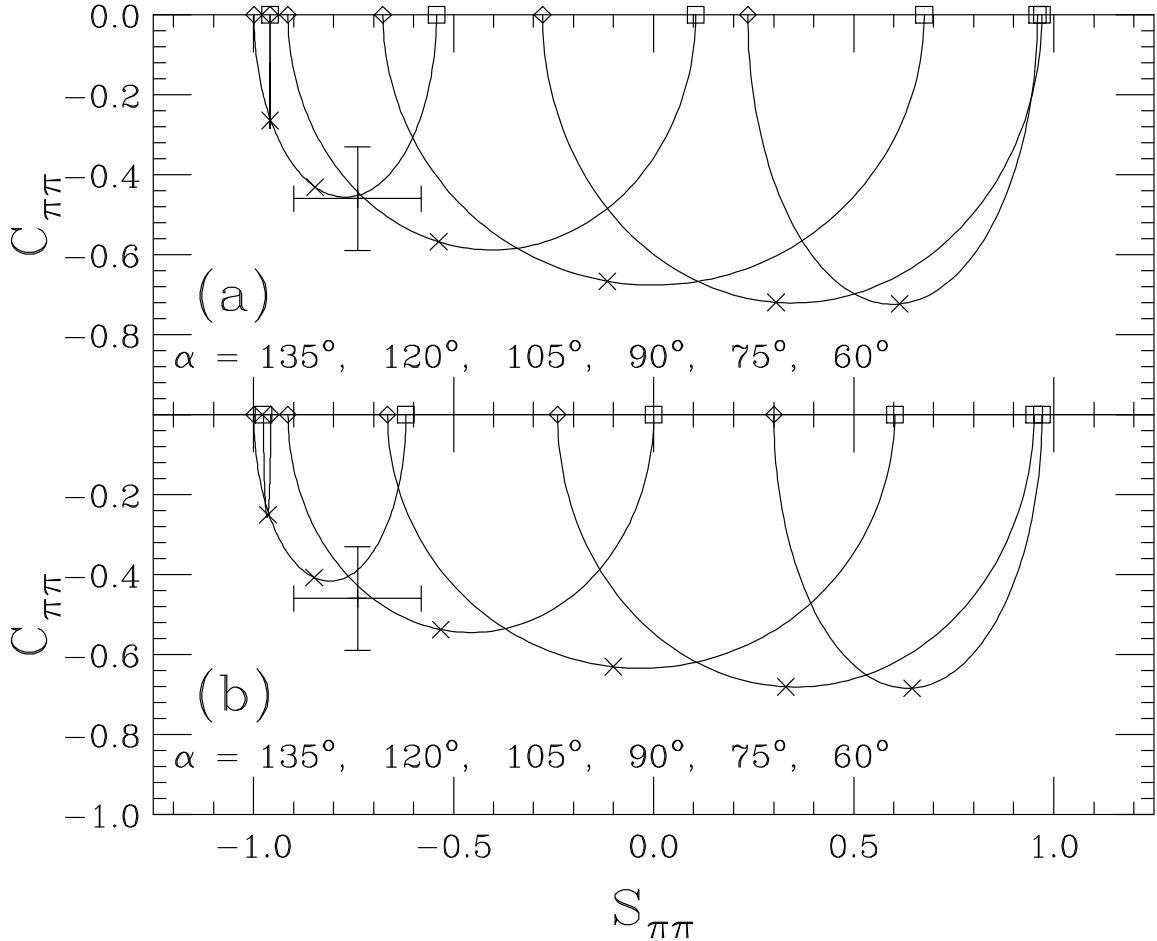}
\caption{Same as Figs.\ 1 but with SU(3) breaking included in penguin terms as
explained in text.}
\end{figure}

The resulting plots of $\cpp$ versus $\spp$ are shown in Figs.\ 2(a) and 2(b).
The constraints on $\alpha$ are stronger than those obtained from Figs.~1(a)
and 1(b) which assumed no SU(3) breaking in $P$.  The use of Fig.\ 2(b) [again,
slightly more restrictive than Fig.\ 2(a)], allows one to exclude values of
$\alpha < 94^\circ$ by the error-ellipse method mentioned in the previous
paragraph.  We would then conclude that $\alpha = (107 \pm 13)^\circ$.  
In any event, this example of SU(3) breaking shows that the limits obtained in
the absence of SU(3) breaking in $P$ are conservative ones.

\bigskip
We thank Andreas H\"ocker, Dan Pirjol, and Jure Zupan for helpful discussions.
JLR is grateful to Maury Tigner for extending the hospitality of the Laboratory
for Elementary-Particle Physics at Cornell during this research.  This work
was supported in part by the United States Department of Energy through Grant
No.\ DE FG02 90ER40560 and in part by the John Simon Guggenheim Memorial
Foundation.

\def \ajp#1#2#3{Am.\ J. Phys.\ {\bf#1}, #2 (#3)}
\def \apny#1#2#3{Ann.\ Phys.\ (N.Y.) {\bf#1}, #2 (#3)}
\def \app#1#2#3{Acta Phys.\ Polonica {\bf#1}, #2 (#3)}
\def \arnps#1#2#3{Ann.\ Rev.\ Nucl.\ Part.\ Sci.\ {\bf#1}, #2 (#3)}
\def \art{and references therein}
\def \cmts#1#2#3{Comments on Nucl.\ Part.\ Phys.\ {\bf#1}, #2 (#3)}
\def \cn{Collaboration}
\def \cp89{{\it CP Violation,} edited by C. Jarlskog (World Scientific,
Singapore, 1989)}
\def \econf#1#2#3{Electronic Conference Proceedings {\bf#1}, #2 (#3)}
\def \efi{Enrico Fermi Institute Report No.}
\def \epjc#1#2#3{Eur.\ Phys.\ J.\ C {\bf#1} (#3) #2}
\def \ib{{\it ibid.}~}
\def \ibj#1#2#3{~{\bf#1} (#3) #2}
\def \ijmpa#1#2#3{Int.\ J.\ Mod.\ Phys.\ A {\bf#1}, #2 (#3)}
\def \ite{{\it et al.}}
\def \jhep#1#2#3{JHEP {\bf#1}, #2 (#3)}
\def \jpb#1#2#3{J.\ Phys.\ B {\bf#1}, #2 (#3)}
\def \kdvs#1#2#3{{Kong.\ Danske Vid.\ Selsk., Matt-fys.\ Medd.} {\bf #1}, No.\
#2 (#3)}
\def \mpla#1#2#3{Mod.\ Phys.\ Lett.\ A {\bf#1}, #2 (#3)}
\def \nat#1#2#3{Nature {\bf#1}, #2 (#3)}
\def \nc#1#2#3{Nuovo Cim.\ {\bf#1}, #2 (#3)}
\def \nima#1#2#3{Nucl.\ Instr.\ Meth.\ A {\bf#1}, #2 (#3)}
\def \npb#1#2#3{Nucl.\ Phys.\ B~{\bf#1} (#3) #2}
\def \npps#1#2#3{Nucl.\ Phys.\ Proc.\ Suppl.\ {\bf#1}, #2 (#3)}
\def \PDG{Particle Data Group, D. E. Groom \ite, \epjc{15}{1}{2000}}
\def \pisma#1#2#3#4{Pis'ma Zh.\ Eksp.\ Teor.\ Fiz.\ {\bf#1}, #2 (#3) [JETP
Lett.\ {\bf#1}, #4 (#3)]}
\def \pl#1#2#3{Phys.\ Lett.\ {\bf#1}, #2 (#3)}
\def \pla#1#2#3{Phys.\ Lett.\ A {\bf#1}, #2 (#3)}
\def \plb#1#2#3{Phys.\ Lett.\ B {\bf#1} (#3) #2} 
\def \prd#1#2#3{Phys.\ Rev.\ D\ {\bf#1} (#3)  #2}
\def \prl#1#2#3{Phys.\ Rev.\ Lett.\ {\bf#1} (#3) #2} 
\def \prp#1#2#3{Phys.\ Rep.\ {\bf#1} (#3) #2}
\def \ptp#1#2#3{Prog.\ Theor.\ Phys.\ {\bf#1}, #2 (#3)}
\def \rmp#1#2#3{Rev.\ Mod.\ Phys.\ {\bf#1}, #2 (#3)}
\def \rp#1{~~~~~\ldots\ldots{\rm rp~}{#1}~~~~~}
\def \yaf#1#2#3#4{Yad.\ Fiz.\ {\bf#1}, #2 (#3) [Sov.\ J.\ Nucl.\ Phys.\
{\bf #1}, #4 (#3)]}
\def \zhetf#1#2#3#4#5#6{Zh.\ Eksp.\ Teor.\ Fiz.\ {\bf #1}, #2 (#3) [Sov.\
Phys.\ - JETP {\bf #4}, #5 (#6)]}
\def \zp#1#2#3{Zeit.\ Phys.\  {\bf#1} (#3) #2}
\def \zpc#1#2#3{Zeit.\ Phys.\ C {\bf#1}, #2 (#3)}
\def \zpd#1#2#3{Zeit.\ Phys.\ D {\bf#1}, #2 (#3)}

\end{document}